# On the Paradox of Chilling Water: Crossover Temperature in the Mpemba Effect

Andrew Wang, Monica Chen, Yanni Vourgourakis and Antonio Nassar

Science Department

Harvard-Westlake School

3700 Coldwater Canyon

Studio City, CA 91604

## I. Introduction

Nothing could seem more mundane and well understood than water. This apparently simple two hydrogen-one oxygen compound is a fundamental component of all living organisms. It covers 70% of our planet and has been extensively studied and has many applications both industrial scientific.[1,2] Yet this extremely common compound has some strange, puzzling and counterintuitive properties. This article is a report on recent efforts to investigate one of the most peculiar of these properties known as the Mpemba effect which states that warm water freezes faster than cold water when they have equal volumes.[3-13]

While there are archives suggesting that ancient scientists dating back to Aristotle have observed the freezing of warmer water faster than cold water, only in recent times has such a phenomenon been experimentally investigated. In 1963, a Tanzanian high-school student (Erasto B. Mpemba) noticed that his hot ice cream mix froze faster than the cold mixes. With great persistence despite the disbelief of his teachers, he insisted on his experimental observations over theory. Since then, a number of experiments have confirmed and reproduced his initial observation.[3-12] Very recent experiments in carbon nanotubes resonators have observed similar unusual properties.[13] Progress in the experimental elucidation of this phenomenon has been hampered largely by the vast number of variables that need to be examined, let alone the possibility of variable dependency. There is no one mechanism that explains the Mpemba effect for all circumstances, but instead, there are different mechanisms important under different conditions.[4] For example, evaporation, dissolved gasses, convection, surroundings and supercooling may all be important to the effect.

Our research focuses not on the point at which water of different temperatures freezes, but rather, the rates of cooling for "warm" and "cold" water. Our report investigates the rates at which hot and cold water cool and the temperature at which "hot" water becomes cooler than cool water after a period of time (this temperature we refer to from here on as the crossover temperature). We have confirmed that warmer water indeed cools at a faster rate than colder water and that, surprisingly, this trend continues past the point where the temperatures of the two samples are the same. Our results show that when using optimal initial temperature conditions, the crossover temperature is found to be 2.7 $^o$C whereas our other set of initial conditions gave a crossover temperature of -0.07 $^o$C. These data taken together provide a definite quantitative evidence of the Mpemba effect.

While most objections to the validity of the phenomenon include the potential significance of environmental factors, we have endeavored to reduce the number of variables to just one by use of vacuums for cooling and double distilled water. Using a vacuum pump to lower the pressure of the atmosphere above the water sample, we were able to cause the liquid to boil. As a result, the most energetic molecules in the water were able to overcome the hydrogen bonds holding them together in liquid phase and escape into the atmosphere. In this way the average temperature of the water will slowly drop as the chamber pressure becomes negative and the molecules with greater kinetic energy become water vapor. This method has the advantage of providing uniform cooling without the possibility of foreign contaminants that may come from external cooling sources such as ice or refrigerants.

## II. Experimental Procedure

To begin the experiment, first set up the apparatus (See Figure 1). To do so, first fill a large beaker with an ice-water mixture. Next, fill one 200 mL beakers with 30 mL of double-distilled water. Drop a magnet stirrer into the beaker and then place the beaker on the hot plate. Set the heat dial to two and the stirring dial to 3. Place an electronic thermometer into the beaker and leave the beaker on the hot plate until it reaches 40 °C. Meanwhile, fill an empty 200 mL beaker with 15 mL of double-distilled water. Place an electronic thermometer into the beaker and then place the beaker into the large beaker filled with ice-water prepared previously for 2 ½ minutes or until the thermometer reads 5 °C.

**Figure 1:**

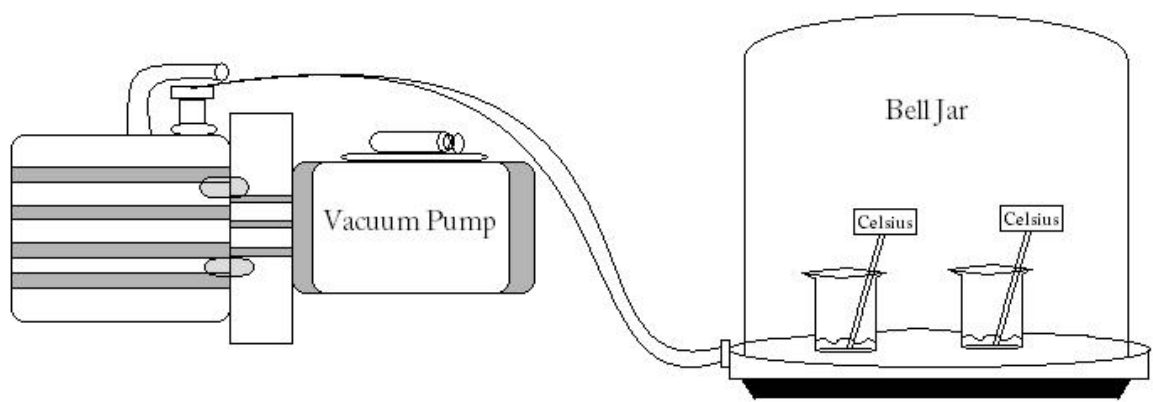

While waiting for the beakers to reach the correct temperatures, fill the vacuum with oil so that it is operable. Then, apply the lubricating grease on the platform of the bell jar. Once the "hot" beaker has reached 40 °C, pour 15 mL of the water into a graduated cylinder. Dump the excess water into the sink and then pour the 15 mL into the empty beaker. After the "cold" beaker has been sufficiently cooled, remove the beaker from the ice-cold mixture and wipe away any ice that might have clung to the beaker.

Place both 200 mL beakers with their respective thermometers into the vacuum and attach the cord connecting the vacuum to the bell jar and put on the bell jar top. When the apparatus is ready and the approximate temperatures are 35 and 5 degrees, turn on the vacuum and start the timer. Take temperature readings for both beakers every 30 seconds for 10-15 minutes. Within the first three minutes, the water should begin boiling. Observe the discrepancy in the boiling rate in the two beakers. The beaker that initially had warmer water should cool faster.

## III. Results

The data collected can be separated into two groups: (1) data collected from optimal initial temperatures suggested in references 7-12 to maximize the Mpemba effect and (2) data collected and averaged from 8 trials with initial temperatures ranging from 1-5 °C (cooler water) and 17-20 °C (warmer water). [Figures 2 and 4, respectively]. There are three important areas of the graphs: the section before the crossover temperature, the crossover temperature and the section after the crossover temperature. In all these sections we find that hot water indeed cools faster than cold water and that for every set of initial water temperatures there is a reproducible crossover temperature unique to that set of conditions.

**Figure 2**: The warmer water is initially slightly above 35 °C while the cooler water is initially at 5 °C.

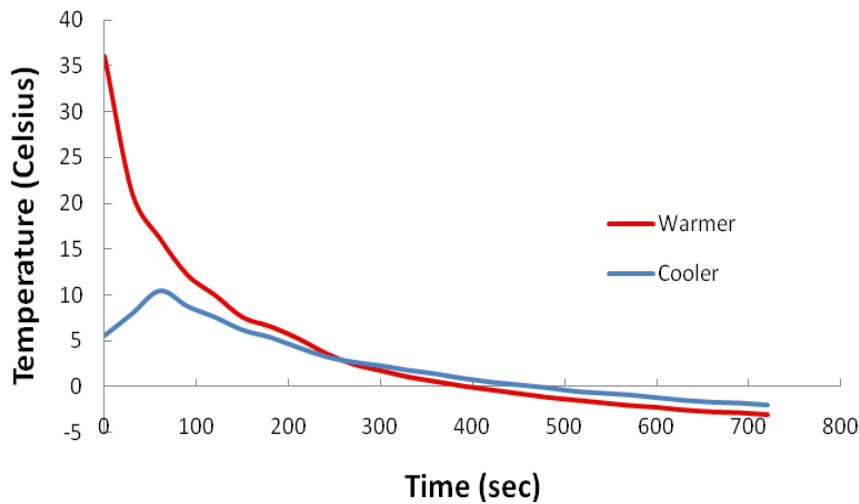

In Figure 2, we use optimum initial temperatures for the Mpemba effect of 35 °C and 5 °C. [7-8] The initially warmer water cools faster than the initially cooler water and this trend continues past the crossover temperature of 2.7 °C. Here, the cold water temperature initially rises as it equilibrates to room temperature and then cools as it begins to boil at about t = 70 s. Interestingly, past the crossover temperature the water initially at higher temperatures continues to cool at a higher rate than the water initially at a lower temperature.

**Figure 3**: Magnification of Figure 2 at the point of crossover.

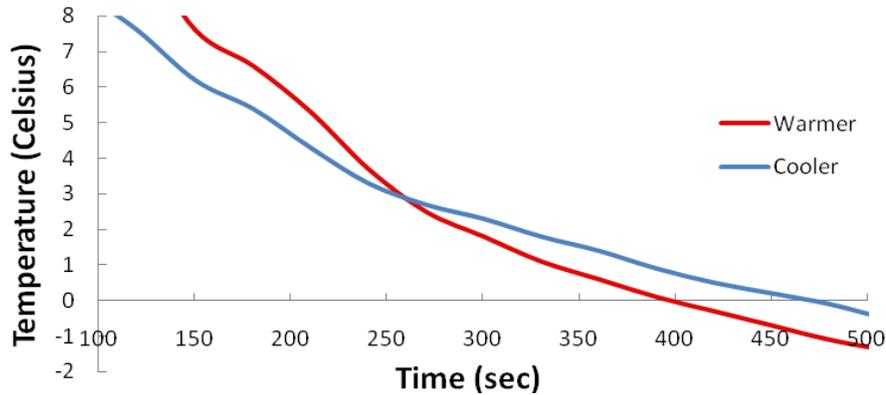

Figure 3 is a close up magnification of Figure 2 and clearly shows that hotter water has more negative slope than cooler water. The crossover temperature for this trial occurs is 2.7 °C. This result was found to be readily reproducible.

**Figure 4**: Here the data were collected and averaged from 8 trials from selected initial temperatures. The initially warmer water temperatures are between 17 and 20 °C and the initially cooler water temperatures are between 1 and 5 °C.

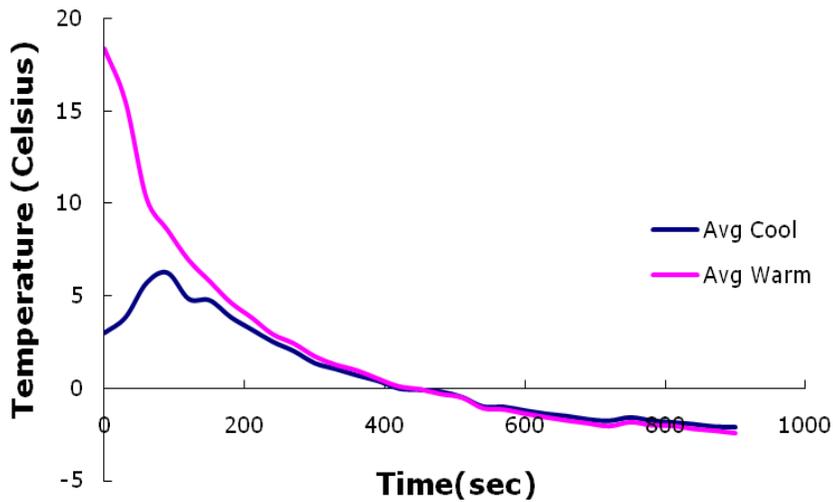

In Figure 4, the data were collected and averaged from 8 trials from selected initial temperatures. The initially warmer water temperatures are between 17 and 20 °C and the initially cooler water temperatures are between 1 and 5 °C. The warm water begins to cool immediately, progressing rapidly after it begins to boil at roughly t = 25 seconds. In comparison, the cold water temperature initially rises as it equilibrates to room temperature and then cools as it begins to boil at roughly t = 90 s. In all trials, the temperature of the two samples converge until the crossover temperature is reached just after 450 seconds. Past the crossover temperature, the water initially at higher temperatures continues to cool at a faster rate than the water initially at a lower temperature.

**Figure 5:** Magnification of Figure 4 showing the region near the crossover temperature.

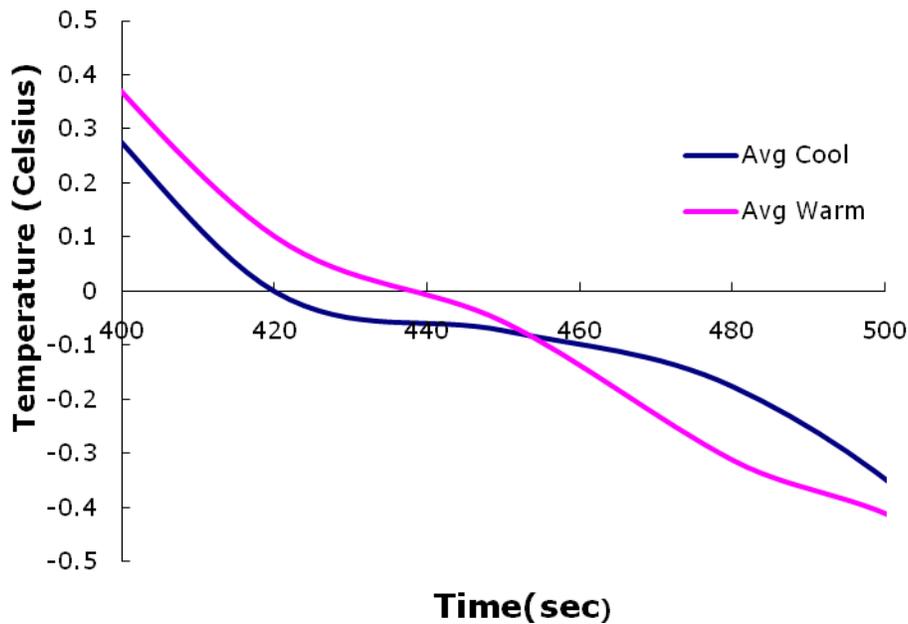

Figure 5 is a close up magnification of Figure 4. It shows clearly that after 450 seconds the crossover temperature occurs at – 0.07 °C. Past this point the initially warmer sample continues to cool at a greater rate than the initially cooler sample. This result was found to be readily reproducible.

## IV. Conclusion

In this paper we have presented data confirming that water initially at higher temperature cools at a faster rate than water initially at a lower temperature and that this trend continues past the point at which the two samples reach the same temperature: *the crossover temperature*. Furthermore, our data indicates that the starting temperature affects the crossover temperature in a reproducible manner. We have confirmed that warmer water indeed cools faster than colder water and that, surprisingly, this trend continues past the point where the temperatures of the two samples are the same. Our results show that when using optimal initial temperature conditions, the crossover temperature is found to be 2.7 $^o$C whereas our other set of initial conditions gave a crossover temperature of -0.07 $^o$C. These data taken together provide a definite quantitative evidence of the Mpemba effect.

Understanding the mechanism of the effect seems a distant goal and while many suggestions have been made none provide a satisfactory explanation of the phenomenon. It has been suggested that mass loss associated with evaporation could result in faster freezing, but this cannot account for demonstrations of the Mpemba effect as a result of dissipative cooling. Others have suggested that since hot water contains less dissolved gases than cooler water that this might be an important factor in this process but there little theoretical or experimental evidence to support this view.

While the existence of the Mpemba effect seems well established now much work needs to be done on order to better characterize and fully understand this effect.
Outstanding questions remain such as what other types of fluids exhibit this effect and the range of temperatures over which the effect can be detected. Is this an effect only associated with water, or other polar liquids, do non-polar liquids have a detectable Mpemba effect? Does this effect work over a wide range of temperatures or is there an upper limit to it? Does the reverse effect exist, where heating rates are affected by initial temperatures? All of these are questions that we hope to address and that will not only widen the spectrum of our inquiry but perhaps suggest avenues through which we can begin to approach a mechanistic understanding of this important phenomenon.

## V. References

1. P. J. Feibelman, "The first wetting layers on a solid", Physics Today 63, 34-39 (2010) and references therein.
2. F. R. Spellman, "The Science of Water: Concepts and Applications", 2$^{nd}$ edition, CRC Press, Boca Raton, FL (2008) and references therein.
3. J. I. Katz, "When hot water freezes before cold" American Journal of Physics 77, 27-29 (2009).
4. M. Jeng, The Mpemba effect: When can hot water freeze faster than cold? American Journal of Physics 74, 514-522 (2006) and references therein.
5. D. Auerbach, "Supercooling and the Mpemba effect: When hot water freezes quicker than cold" American Journal of Physics 63, 882-885 (1995) and references therein.
6. G. S. Kell, "The freezing of hot and cold water," American Journal of Physics 37(5), 564–565 (1969).
7. R. M. Robson, "Mpemba's ice cream," New Scientist 43, 89 (1969).
8. M. B. F. Ranken, "Mpemba explained," New Scientist 45, 225–226 (1970).
9. R. Jephson, "Mpemba's ice cream," New Scientist 42, 656 (1969).
10. J. C. Dixon, "Mpemba's ice cream," New Scientist 42, 656 (1969).
11. F. L. C. Blackwall, "Mpemba's ice cream," New Scientist 43, 88–89 (1969).
12. S. M. Arkless, "Mpemba's ice cream," New Scientist 42, 655–656 (1969).
13. P. A. Greaney, G. Lani, G. Cicero and J. C. Grossman, "Anomalous Dissipation in Single-Walled Carbon Nanotube Resonators", Nano Letters 9, 3699-3703 (2009) and references therein.